\newcommand{\psipdc}{\psi_{\rm PDC}}   		
\newcommand{\psisfg}{\psi_{\rm SFG}}   		
\newcommand{\Fpdc}{F_{\rm PDC}}  			
\newcommand{\Fsfg}{F_{\rm SFG}}                   
\newcommand{\w}{\vec{w}}			

\newcommand{\vxi}{\vec{\xi}}				

\newcommand{\x}{\vec{x}}

\newcommand{\sinc}{{\rm Sinc}}
\newcommand{\q}{\vec{q}}

\newcommand{\bsub}{\begin{subequations}}
\newcommand{\esub}{\end{subequations}}
\newcommand{\beq}{\begin{equation}}
\newcommand{\eeq}{\end{equation}}
\newcommand{\beqa}{\begin{eqnarray}}
\newcommand{\eeqa}{\end{eqnarray}}
\newcommand{\beql}{\begin{subequations}\begin{eqnarray}}
\newcommand{\eeql}{\end{eqnarray}\end{subequations}}
\documentclass[aps,prl, groupedaddress,amsmath,showpacs]{revtex4}
\usepackage{amsmath}
\usepackage{graphicx}
\usepackage{float}   
\usepackage{verbatim}  
\usepackage{braket}
\begin{document}
\title{Detection of  the ultranarrow  temporal correlation \\ of twin beams via sum-frequency generation}
\author{O. Jedrkiewicz$^{1,2}$, J.-L. Blanchet$^1$, E.~Brambilla$^1$,  P. Di Trapani$^1$  and A.~Gatti$^{1,2}$\footnote{Corresponding Author: Alessandra.gatti@mi.infn.it}}
\affiliation{$^1$CNISM and Dipartimento di Alta Tecnologia Universit\`a dell'Insubria, Via Valleggio 11 Como, Italy,
$^2$CNR, Istituto di Fotonica e Nanotecnologie, Piazza L. da Vinci 4 Milano, Italy}
\begin{abstract}
 We demonstrate the ultranarrow temporal correlation (6 fs full width half maximum) of twin beams generated by  parametric down-conversion, by using the inverse process of sum-frequency generation. The result relies on an achromatic imaging of a huge bandwith of twin beams and on a careful control of their spatial degrees of freedom. The detrimental effects of spatial filtering and of imperfect imaging are shown toghether with the theoretical model used to describe the results. 
\end{abstract}
\pacs{42.50.-p,42.50.Dv, 42.65.Lm	}
\maketitle

The entangled photon pairs produced by parametric down-conversion (PDC)  are the key elements for several quantum communication and metrology schemes. Crucial to these applications is the ability of tailoring their temporal correlation properties. In particular, various methods of generating ultra-broadband biphotons, based e.g on the engineering of the nonlinear medium \cite{Nasr2008} or of the pump \cite{Donnel2007},  have been recently developed.
\par 
 Our approach relies rather on the peculiar X-shaped geometry of the spatio temporal correlation of  twin photons, theoretically outlined by some of us \cite{Gatti2009,Caspani2010}.  These investigations predicted the possibility of manipulating the temporal bandwidth of PDC entanglement by acting on the spatial degrees of freedom, and,  in particular,  of achieving an anusual  relative temporal localizations of twin photons (few femtoseconds) when their near-field positions are resolved.

In this work we report  the experimental observation of such ultra-narrow temporal correlation  (6 fs full width half maximum )  of twin beaams,  detected by means of the inverse process of sum-frequency generation (SFG). The temporal correlation profile  is measured by introducing a controlled temporal delay between twin beams, and then imaging the output of the PDC crystal  onto the input face of a second crystal where up-conversion takes place.

SFG  is used in classical optics as an ultrafast correlator. In quantum optics, recent experiments \cite{Dayan2005, Donnel2009, Sensarn2010} used SFG to probe the temporal correlation of twin photons.
Contrary to the Hong-Ou-Mandel scheme, which is insensitive to dispersion (see e.g. the results reported in \cite{Nasr2008}), in the SFG scheme the presence of dispersive optical elements drastically broadens the measured temporal correlation of biphotons.
Indeed, the experiments \cite{Dayan2005, Donnel2009} used prisms to correct the dispersion introduced by  optical lenses, which were effective over a large but still limited bandwidth of PDC. As a result the measured temporal correlation of biphotons was in the range  $\approx 30$  to $ 100$ fs.
\par
Our setup (Fig.\ref{fig_1})  presents a number of distinguishing features:
\begin{figure}[h!]
\includegraphics[width=10.5cm,keepaspectratio=true]{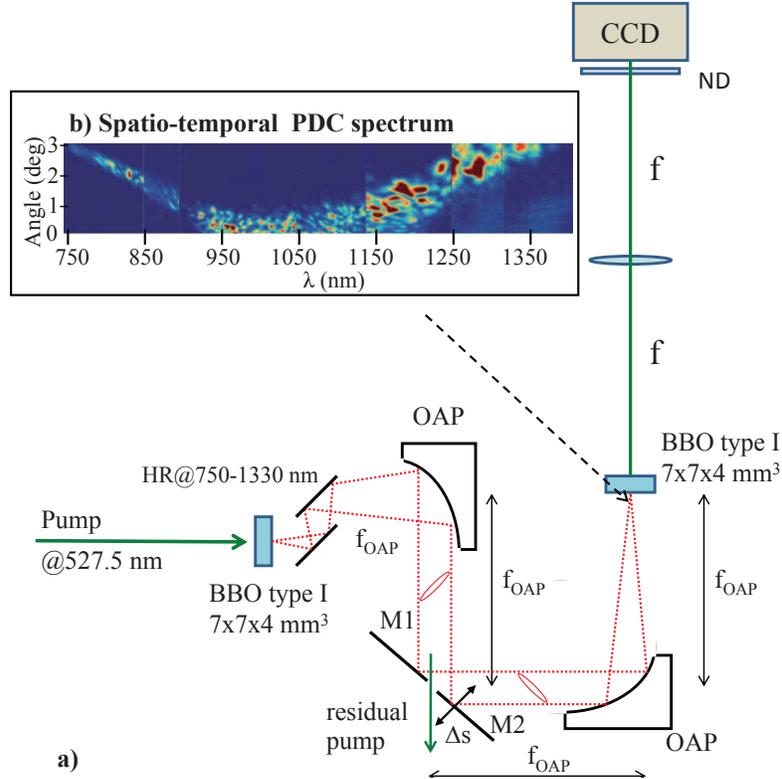}
\caption{(Color online)(a)Experimental layout for the observation of the twin beam temporal correlation b) Spatio-temporal spectrum of the PDC light measured by an imaging spectrometer just before the up-converting crystal. Note the huge spatio-temporal bandwidths reflected by the OAP mirrors (the colourmap does not reflect the real intensity, different spectral portion having being detected with different attenuations)}
\label{fig_1}
\end{figure}
i) rather than correcting the temporal dispersion introduced by lenses, we implement an achromatic imaging of the PDC light onto the SFG crystal by using parabolic mirrors; 
ii) a broad spatial and temporal bandwidth of the PDC light is imaged (in phase and amplitude) onto the SFG crystal. The non-factorability of the correlation in space and time \cite{Gatti2009,Caspani2010} implies that both collecting a huge spatial bandwidth and performing a perfect spatial imaging are  key elements to preserve the ultra-narrow temporal localization of twin photons\cite{Brambilla2011}. Indeed our data will show that  the effects of diffraction in free space are detrimental to the temporal localisation in a way similar to dispersion. 
These features allow to preserve the phase conjugation of twin photons over a huge bandwidth ($\approx 600$ nm), enabling thus the demonstration of their ultra-narrow temporal localization, the narrowest -to the best of our knowledge- measured in experiments using the SFG process as a probe for PDC correlation.
\par
The experimental layout is shown in Fig.\ref{fig_1}a. The pump pulse at $\lambda_0 =527.5$nm is obtained from the second harmonic of a 1 ps, 1055 nm, 10 Hz repetition rate Nd:Glass laser (Twinkle, Light Conversion Ltd.), and is collimated down to about 0.8mm (FWHM) at the entrance of a type I 4mm Beta Barium Borate (BBO) crystal for PDC in the collinear configuration. 
Just after the crystal, two custom made high reflectivity dielectric mirrors (Layertec),  with reflectivity $R >99.9 \%$ in the 750-1330 nm range and $R\approx 0,3 \%$ at $\lambda_0=527.5$nm, are used to reflect the PDC radiation. The achromatic imaging is performed by means of two identical 90$^0$ off-axis parabolic gold mirrors (OAP), which reflect a huge portion of the spatio-temporal spectrum the PDC radiation, as shown in Fig.\ref{fig_1}b, and image  the output plane of the PDC crystal onto the entrance face of  a second BBO crystal, identical  to the first one, placed at the plane $4f_{\mathrm OAP}$. 
Two adjacent gold mirrors M1 and M2 are placed in the far-field plane of the PDC source at $2f_{\mathrm OAP}$. They act separately on the two twin-beam components of the light, because each photon has its twin on the opposite side of the far-field plane due to momentum conservation in the elementary PDC process.  Mirror $M_2$ can be translated to produce a relative delay between twin beams. A 1.5mm wide gap between the two mirrors allows to eliminate the residual input pump. The SFG crystal is placed in the focal plane of the second OAP mirror and is mounted on a micrometer translation stage permitting to finely adjust its position relative to the imaging plane $4f_{\mathrm OAP}$.  Both crystals are also mounted on a rotation stage in order to adjust their orientation with the aim of working at exactly the same phase-matching conditions.
The far-field radiation emitted by the SFG crystal is observed in the focal plane of a 20cm focal length lens and the light intensity is monitored by means of a 16 bit scientific CCD camera (Roper Scientific) with 80 $\%$ detection efficiency at 527.5 nm. The far-field SFG distribution shows a  narrow central peak, representing the coherently reconstructed  far-field profile of the original pump, lying over a widely spread speckled background (see \cite{Jedr2011}). The latter originates from  incoherent SFG processes as well as from the residual PDC not up-converted. Note that PDC light is  absent  at the location of the coherent peak, because the central portion of the PDC far-field  is eliminated by the gap between $M_1$  and $M_2$.  
\par
\begin{figure}[h!]
\includegraphics[width=8cm,keepaspectratio=true]{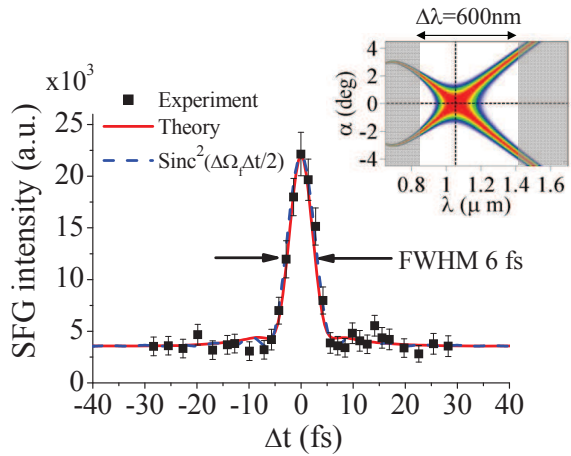}
\caption{(Color online) Temporal correlation profile of twin beams produced by a type I BBO as measured by monitoring the SFG peak intensity as a function of the temporal delay between the twin components. Superimposed to experimental data the red line shows our theoretical calculation. The blue dashed line is   $ \sinc^2 (\Delta \Omega_{f} \Delta t /2)$ . }
\label{fig_2}
\end{figure}

Fig.\ref{fig_2} reports our main result, and shows the SFG peak intensity monitored in the central pixel of the far-field distribution as a function of the temporal delay, and for an optimum position of the second BBO crystal along the propagation direction. Each data point corresponds to the coherent peak intensity averaged over 15 images, each of it recorded over 2s (20 laser shots). The  lines superimposed to the experimental data show our theoretical predictions, according to  the model  described in the following. 

The scheme is modelled semi-analytically, exploiting the monochromatic and plane-wave pump (PWP) approximation. This model is elaborated in details in \cite{Brambilla2011}, while in the following we report only the key results.

The main quantity of interest, characterising the entanglement of twin beams generated by PDC is the {\em biphoton correlation}
(or {\em biphoton amplitude} )
\beq
\psipdc= \langle A (\xi) A(\xi + \Delta \vxi) \rangle
\label{psidef}
\eeq
where $A$ is the field operator of the down-converted signal field and $\vxi =(\x,t)$ is the 3-D spatio-temporal coordinate,  $\x$ denoting the transverse position in the beam cross-section at the output plane of the crystal.

In the PWP limit, the equation describing the generation of PDC light in the nonlinear crystal can be exactly solved (see \cite{Caspani2010, Brambilla2011}). In the same limit, the biphoton correlation depends only on the spatio-temporal separation $\Delta \vxi$. It can be written as the spatio-temporal Fourier transform of the probability amplitude
$ \Fpdc(\w)$ of generating a photon pair in the symmetric Fourier modes $\w$ and $-\w$, where $\w= (\q, \Omega)$,  $\q$ being the transverse wave-wector and $\Omega$ the offset from the central frequency $\omega_1=\omega_0/2$:
\beq
\psipdc (\Delta \vxi) = \int \frac{d \w} {(2\pi)^3} e^{i \w \cdot \Delta \vxi}  \Fpdc (\w) \, .
\label{psi}
\eeq
The probability amplitude $\Fpdc$ is strongly peaked around the curve where phase matching occurs, i.e, where
\beq
\Delta (\w) = l_c \left[ k_{1z} (\w) + k_{1z}(-w) -k_0\right]  =0 \, ,
\label{PM}
\eeq
with $k_{1z} (\w) = \sqrt{ k_1 (\Omega) ^2 -q^2}$ being the z-component of the signal wave-vector,  $k_0$  the pump wave-number, and $l_c$ the crystal length.    For example, in the low-gain limit $ g \ll 1$,  it takes the well known form
$\Fpdc (\w) \approx  g \, \sinc \left[  \frac{\Delta (\w)}{2}\right] \exp{( i \Delta (\w)/2)} $,
where $g$ is a  gain parameter proportional to the pump amplitude and to the nonlinear susceptibility (the general form of $\Fpdc$ can be found e.g. in \cite{Brambilla2011}).
For collinear phase matching, i.e. for $2 k_1 (0)= k_0 $, the region where $\Fpdc$ takes its maximal value follows
a characteristic hyperbolic geometry in the plane $ (|q|, \Omega) $, as the spatiotemporal spectrum shown in Fig. \ref{fig_1}b. Indeed, a quadratic expansion of Eq. \eqref{PM}, valid close to degeneracy gives :
\beq
\Delta (\q, \Omega) \approx  -\frac{q^2 \, l_c}{ k_1} + k_1^{''} \, l_c \Omega^2  \, .
\label{PM2}
\eeq
Thus, phase matched modes are those for which the temporal dispersion occuring along the crystal is compensated by diffraction. Because of this geometry of phase matching, the spatio-temporal correlation \eqref{psidef})  assumes a X-shape
in any plane containing time and a spatial coordinate. A key result of \cite{Gatti2009, Caspani2010} is that when twin photons are detected at the same near-field position the width of their temporal correlation, i.e the width of $\psipdc(\Delta x=0, \Delta t)$,  is given by the inverse of the {\em full temporal bandwidth} detected (that can extend in principle to the pump optical frequency),  and not by the GVD bandwidth $1/\sqrt{k"_1 l_c}$ characterizing the far-field correlation. This is a consequence of the cancelation of temporal dispersion by diffraction occurring for all the generated modes[see Eq.\eqref{PM2}]. Therefore, a temporal correlation  in the femtosecond range can be in principle achieved, provided that twin photons are localized in the near-field.
\par
The SFG propagation equation in the second crystal can be solved in the limit where the fraction of the light that is up-converted is small.
Our calculations in \cite{Brambilla2011} show that at the crystal output the SFG intensity  distribution is homogeneous in space and time, (a mere consequence of the PWP approximation), and has two components:
$I_{\rm SFG}= I_{\rm SFG}^{\rm incoh} + I_{\rm SFG}^{\rm coh}$. The incoherent component results from the random up-conversion of photons originally unpaired. All the information on the twin-beam correlation is contained in the coherent component, which originates  from the up-conversion of phase conjugate photons \footnote{In the high gain regime the coherent contribution originates not only from photons which were originally produced as twins, but from all those signal-idler pairs that are correlated because of cascading PDC processes}, and has the form \cite{Brambilla2011}:
\beq
I_{\rm SFG} ^{\rm coh} = \left| \int \frac{d\w} {(2\pi)^3} H_+ (\w) H_-(-w) \Fpdc (\w) \Fsfg(-\w) \right|^2
\label{Icoh}
\eeq
where $\Fsfg= \sigma l_c^\prime \sinc \Delta_{\rm SFG} (\w) \exp{i \Delta_{\rm SFG} (\w) /2}$ represents the probability amplitude of up-converting a pair of photons in modes $\w$ and $-\w$, which depends on the phase matching $\Delta_{\rm SFG} (\w) $ in the second crystal.  $H_+$ and $H_-$ are optical transfer functions describing the propagation between the two crystals
of the twin components of the PDC light  with  $q_x>0$ and $q_x<0$, respectively, separated in the far-field by mirrors $M_1$ and $M_2$.
 Expression \eqref{Icoh} tells us that the coherent SFG component results from the sum over all the probability amplitudes of a photon pair being generated in the first crystal  in modes $\w,\,  -\w$ times the probability amplitude the same photon pair is upconverted in the second crystal.
\par
Assuming that a small  temporal delay $\Delta t$ is introduced by mirror $M_2$,  in the ideal case of perfect imaging, with no   dispersive optical elements or losses, we have $H_+(\w)=1$,  $H_-(\w) = e^{i \Omega \Delta t }$, and Eq.\eqref{Icoh} becomes:
\beqa
I_{\rm SFG} ^{\rm coh} &=& \left| \int \frac{d \q d \Omega} {(2\pi)^3} e^{-i \Omega \Delta t} \Fpdc (\q, \Omega) \Fsfg(-\q, -\Omega) \right|^2
\label{Ideal1} \\
&=& \left| \left[ \psipdc \otimes \psisfg\right (\Delta \x =0, \Delta t)  \right|^2
\label{Ideal2}
\eeqa
where $\psisfg (\Delta \vxi) = \int {d \w}/ (2\pi)^3  e^{i \w \cdot \Delta \vxi}  \Fsfg (\w) $ is the biphoton amplitude in the second SFG crystal, and in writing Eq.\eqref{Ideal2} we used the convolution theorem toghether with Eq.\eqref{psi}. In the limit of a  short SFG crystal, $\psisfg (\Delta \vxi) $ behaves as a Dirac-delta in the convolution \eqref{Ideal2}, so that  the PDC correlation can be exactly reconstructed by monitoring the SFG intensity as a function of $\Delta t$. However, it is possible to reconstruct the  shape and width of the temporal correlation also for a finite SFG crystal, provided that it is  tuned exactly  for the same phase matching conditions as the PDC crystal. In this case, the probability amplitudes $\Fpdc$ and $\Fsfg$ for down-conversion and upconversion overlap in the $(\q, \Omega)$ space, so that the integral in Eq. \eqref{Ideal1} well reproduces $|\psipdc (\Delta x=0, \Delta t)|^2$. The red line in Fig. \ref{fig_2} has been plotted from Eq.\eqref{Ideal1}, where the integration was limited to a bandwidth $\Delta \Omega_{f} = 0.9 \times 10^{15} $Hz simulating the finite transmission bandwidth of the setup, and the baseline of the curve was adjusted to account for the experimental background (incoherent SFG + scattering + residual PDC light). 
Remarkable is that both the theoretical and measured profiles are well fitted by the simple curve  $ \sinc^2 (\Delta \Omega_{f} \Delta t /2)$, which is just the Fourier transform of a box function of width $\Omega_f$ in the frequency domain. This gives clear evidence to our claim that dispersion is canceled by diffraction for the transmitted spectral modes, so that they all contribute to a localized temporal correlation peak summing up coherently.
\par
\begin{figure}[t]
\includegraphics[width=7.2cm,keepaspectratio=true]{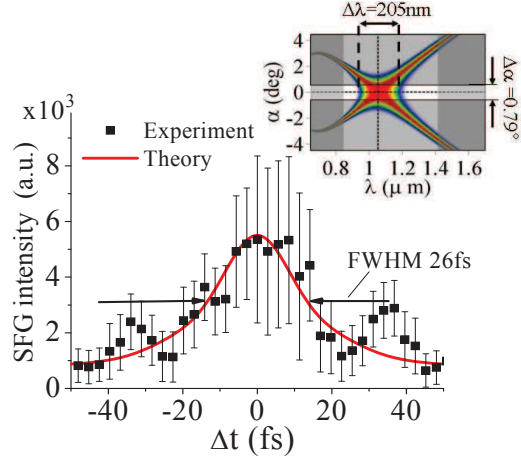}
\caption{(Color online)Effect of spatial filtering. A pinhole clips an angular sector $\Delta \alpha= 0.79^0$ of the PDC light(see the inset). The reconstructed temporal correlation profile broadens, as a consequence of the reduction of the effective temporal bandwidth of phase matching.}
\label{fig_clip}
\end{figure}

The measurement has been repeated in non ideal conditions. Fig.\ref{fig_clip} shows the effect of a spatial filtering of twin beams, performed by inserting in the far field of the PDC crystal  a circular pinhole, which clips an angular portion of the conical emitted radiation. In particular,  the temporal correlation profile plotted in Fig.\ref{fig_clip} was obtained by placing a 4mm diameter pinhole at 29 cm from the output face of the crystal. The data show a remarkable broadening of the {\em temporal} correlation, when filtering is performed over {\em spatial} degrees of freedom, which represents a first proof of principle demonstration of the non-factorable character of the PDC correlation \cite{Gatti2009}. The broadening of the temporal peak occurs because,  by cutting spatial modes, phase matching occurs inside a smaller temporal bandwidth $\Delta \Omega_{eff} \approx 0.34 \times  10^{15} Hz$, as schematically shown by the inset of Fig.\ref{fig_clip}.
\par
\begin{figure}
\includegraphics[width=10cm,keepaspectratio=true]{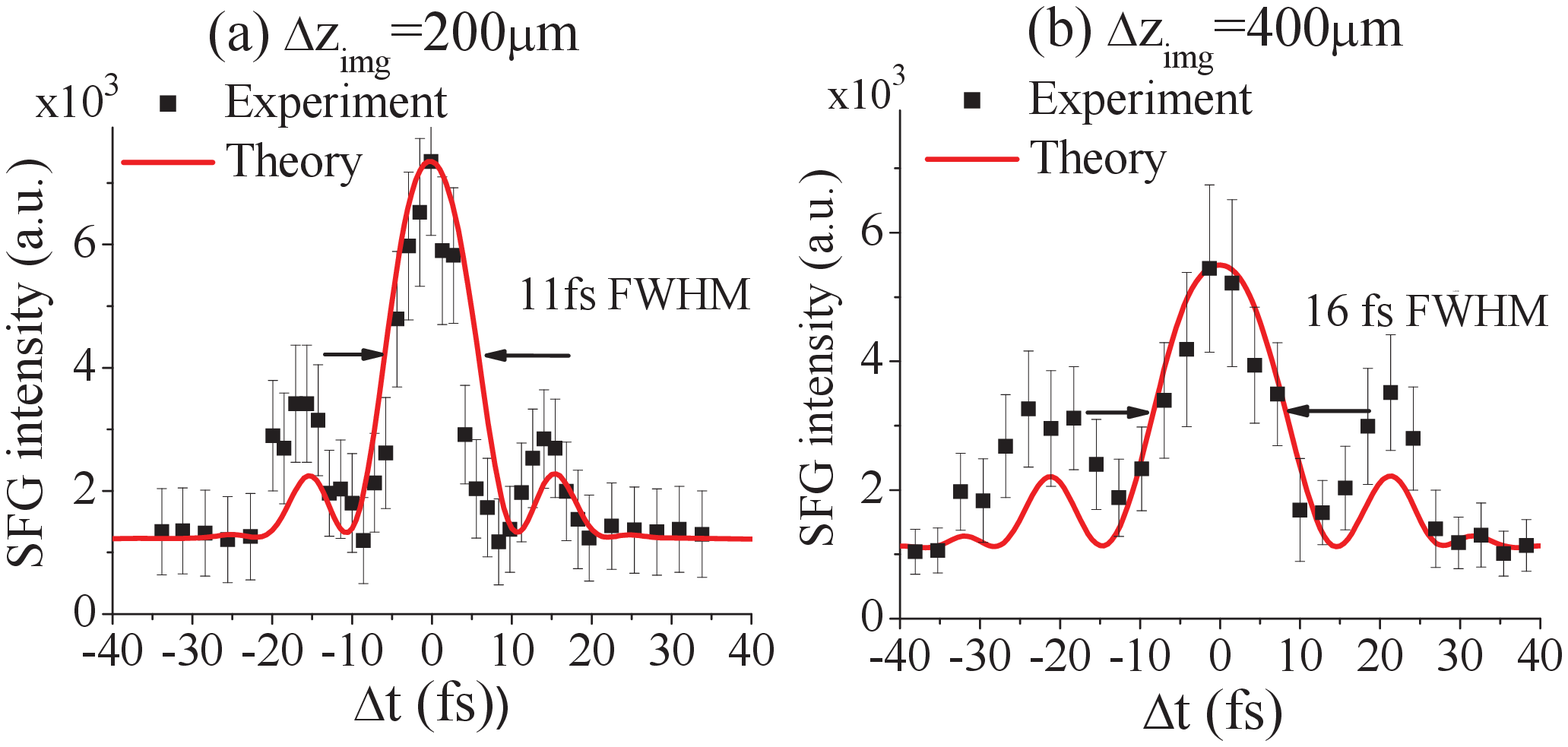}
\caption{(Color online) Effect of an imaging error on the reconstructed temporal corrrelation profile. When the second crystal is traslated by $\Delta z_{ \rm img}$ with respect to the $4f_{\mathrm OAP}$
 plane, the reconstructed temporal correlation broadens.}
\label{fig_imag}
\end{figure}
Even more remarkable are the effects of an imperfect imaging of the PDC light onto the SFG crystal.  We repeated the measurement
by displacing the SFG crystal by an amount $\Delta z_{img}$ away from the optimal imaging plane at $4f_{\mathrm OAP}$. The results, showing how the temporal correlation progressively undergoes broadening when an imaging error is introduced in the set-up, are presented in Fig.\ref{fig_imag}. 
These data show how the diffraction introduced by small displacements (few hundreds microns)   with respect to the optimal imaging plane i) reduces drastically the efficiency of up-conversion,  and ii) broadens the temporal correlation. The second effect is not trivial, and again shows the inderdependence of spatial and temporal degrees of freedom, because spatial diffraction in free space has an effect on the temporal correlation similar to temporal dispersion: an error  in the imaging plane introduces a propagation phase that prevents the spectral modes in the bandwidth from summing up coherently in a narrow temporal peak. Formally, when the SFG crystal is shifted by  $\Delta z_{img}$ from the $4f_{\mathrm OAP}$ plane,  an additional propagation  phase is introduced under the integral in Eq. \eqref{Icoh}, so that
$H_+(\w)H_-(-\w) =  e^{-i \Omega \Delta t} \exp{[-i \frac{q^2 c}{\omega_1 (1-\Omega^2/\omega_1^2)}  \Delta z_{img} ]} $ \cite{Brambilla2011}. If we now make the simple assumption that only phase matched modes contribute to the coherent SFG peak, we can use Eq.\eqref{PM} to substitute $q^2 \to \Omega^2 k_1 k_1^{''}  $, with $k_1= n_1(\omega_1) \omega_1/c$.   The transfer function thus becomes: 
$H_+(\w)H_-(-\w) \to  e^{-i \Omega \Delta t} \exp{[-i \frac{\Omega^2  k_1^{''}  n_1(\omega_1)  }{ 1-\Omega^2/\omega_1^2}  \Delta z_{img} ]} $ , which describes a quadratic dispersion-like chirp of the twin beams. 
\par
In conclusions, we have been able to demonstrate- for the first time to our knowledge- that  the temporal correlation of twin beams is as narrow as few femtosecond,  in a setup that uses the inverse SFG process. This result relies not only on minimizing the temporal dispersion,  thanks to an achromatic imaging, but also on the control of the spatial degrees of freedom of twin beams. Our counterexamples of Fig.\ref{fig_clip} and \ref{fig_imag}  indeed show that spatial filtering and free-space diffraction  broaden the temporal correlation, giving thus evidence of the interdependence of spatial and temporal degrees of freedom in  PDC claimed by theory \cite{Gatti2009,Caspani2010, Brambilla2011}. 

\acknowledgments{We acknowledge support from the grant 221906 HIDEAS of the Fet Open Programme of the EC}

\end{document}